\documentclass[aps,prapplied,reprint]{revtex4-2}

\usepackage{etoolbox}
\usepackage{hyperref}
\usepackage{amsmath}
\usepackage{amssymb}
\usepackage{dsfont}
\usepackage[T1]{fontenc}
\usepackage{ae}
\usepackage{color}
\usepackage{graphicx} 
\usepackage{placeins}

\usepackage[mediumspace,mediumqspace,squaren]{SIunits} 

\begin{document}

\title{ High-Resolution Imaging of Cold Atoms through a Multimode Fiber}

\author{Nicolas Vitrant} \affiliation{JEIP,  USR 3573 CNRS, Coll{\`e}ge de France, PSL University, 11, place Marcelin Berthelot, 75231 Paris Cedex 05, France}

\author{S{\'e}bastien Garcia} \affiliation{JEIP,  USR 3573 CNRS, Coll{\`e}ge de France, PSL University, 11, place Marcelin Berthelot, 75231 Paris Cedex 05, France}

\author{Kilian M{\"u}ller}  \email{Present address: LightOn, 3 impasse Reille, 75014 Paris, France} \affiliation{JEIP,  USR 3573 CNRS, Coll{\`e}ge de France, PSL University, 11, place Marcelin Berthelot, 75231 Paris Cedex 05, France}

\author{Alexei Ourjoumtsev} \email{Corresponding author: alexei.ourjoumtsev@college-de-france.fr} \affiliation{JEIP,  USR 3573 CNRS, Coll{\`e}ge de France, PSL University, 11, place Marcelin Berthelot, 75231 Paris Cedex 05, France}

\begin{abstract}
We developed an ultra-compact high-resolution imaging system for cold atoms. Its only in-vacuum element is a multimode optical fiber with a diameter of \unit{230}{\micro\meter}, which simultaneously collects light and guides it out of the vacuum chamber. External adaptive optics allow us to image cold Rb atoms with a $\sim$ \unit{1}{\micro\meter} resolution over a \unit{100 \times 100}{\micro\meter\squared} field of view. These optics can be easily rearranged to switch between fast absorption imaging and high-sensitivity fluorescence imaging. This system is particularly suited for hybrid quantum engineering platforms where cold atoms are combined with optical cavities, superconducting circuits or optomechanical devices restricting the optical access.
\end{abstract}

\maketitle

Cold atoms are powerful resources for quantum engineering. Controlling them quickly and precisely requires high-resolution optics compatible with an ultra-high-vacuum (UHV) environment. The development of optical systems designed to manipulate single atoms~\cite{Schlosser2001,Barredo2018} and of ``quantum gas microscopes'' capable of resolving individual atoms trapped in optical lattices \cite{Bakr2009,Sherson2010} has led to major breakthroughs, in particular for quantum simulations. Things become more difficult in hybrid setups, combining atoms with optical cavities, superconducting circuits or nanostructures, where mechanical constrains prevent the use of bulk high-resolution optics. Imaging cold atoms with a high resolution and guiding the light out of the vacuum system are two seemingly simple tasks which, in practice, limit the performance of many experiments.

Optical fibers intuitively seem to be a good solution for both problems at once. Single-mode fibers have indeed been used to trap cold atoms in pre-determined configurations \cite{Vetsch2010,Garcia2013} but, by definition, they cannot transmit a multi-mode image. As for multi-mode fibers, they were traditionally disregarded, as they tend to completely randomize the transmitted wavefronts.
A paradigm shift occurred with the recent development of adaptive optics. Even though a sufficiently long multimode fiber is indeed a random optical medium, it is a linear one. The transmission matrix relating the light fields in a set of input and output modes can be measured \cite{Popoff2010}. Its inverse transforms numerically randomly-looking images registered at the ``proximal'' end of the fiber into meaningful images of an object located near the ``distal'' end. Alternatively, the wavefronts emerging from the fiber can be optically reshaped using spatial light modulators. 

In this paper, we extend these two methods, initially developed in the context of biological imaging \cite{Papadopoulos2013}, to the imaging of cold atoms.
We use a highly multimode optical fiber, acting both as a collection optic and a waveguide, to image $^{87}$Rb atoms optically trapped near its distal end. The fiber's total diameter of \unit{230}{\micro\meter} is small enough for it to be used in many spatially-constrained setups. Its high numerical aperture allows us to reach a \unit{1.2}{\micro\meter} resolution in a \unit{100 \times 100}{\micro\meter\squared} field of view. We start by describing the experimental apparatus and the way we measure the fiber's transmission matrix. We then present our results on absorption imaging of atomic clouds by digital inversion of the transmission matrix. The shadow cast by the atoms on a coherent beam injected in the fiber is reconstructed from the speckle patterns of the emerging light. We also introduce an alternative, scanning-point technique, allowing for both absorption imaging and incoherent fluorescence detection of small atomic samples. The light emitted from a point at the distal end of the fiber is refocused at the proximal end by a spatial light modulator, thus performing an optical inversion. 

\begin{figure}[t]
\centering
\includegraphics[width=85mm]{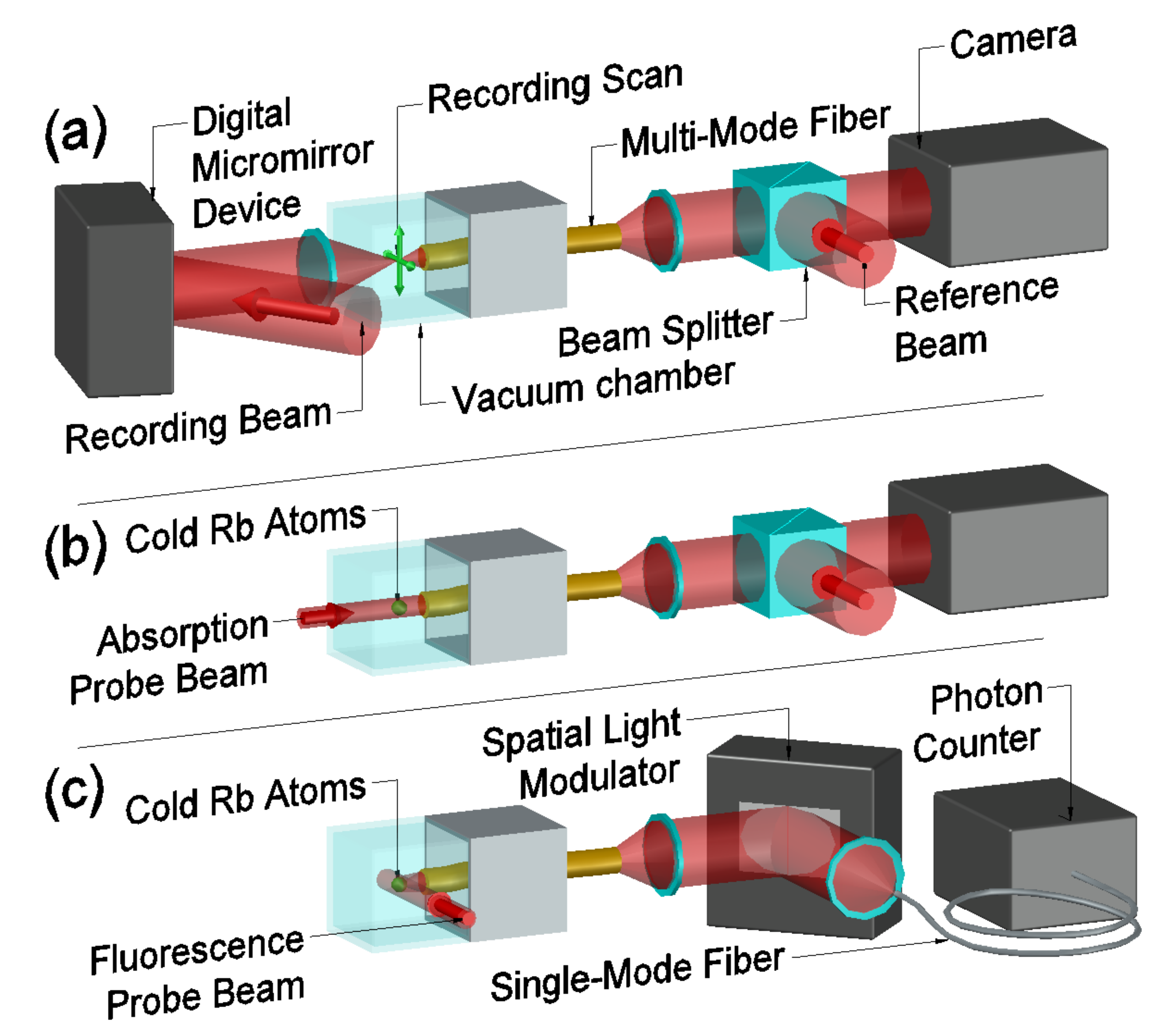}
\caption{Experimental setup used for imaging of cold atoms through a multimode fiber in different configurations. (a) Characterization of the transmission matrix of the fiber: we change the input mode with a digital micromirror device (DMD) while interferometrically recording the output field on a camera. (b) Absorption imaging by digital inversion: we use the inverted transmission matrix to see the shadow of the atoms on a laser coupled in the fiber. (c) Fluorescence detection by optical inversion: we couple the atom emission mode to a photon counter with a spatial light modulator (SLM).}
\label{fig:setup}
\end{figure}

In the experimental setup, we use a $25$-cm-long step-index multimode fiber (MMF), with a core diameter of \unit{200}{\micro\meter} and a numerical aperture of $0.5$ (Thorlabs FP200ERT), as a compact optical bridge between the inside and the outside of an ultra-high-vacuum (UHV) chamber, see Fig.~\ref{fig:setup}. Inside a glass cell of the UHV chamber, we create a cloud of cold rubidium 87 atoms loaded from background vapor in a magneto-optical trap, located $3\,$mm from the in-vacuum distal end of the MMF. As described in Ref.~\cite{Vitrant2020}, we use an $800$-nm laser to transport cold atoms to \unit{200}{\micro\meter} from the MMF distal end and to store them there in a micro-trap. This dipole-trapping laser is shaped by digital optical phase conjugation (DOPC) and sent through the MMF from the outside-vacuum proximal end. On this side, a dichroic mirror allows to separate the $800$-nm-wavelength light from the one of the $780$-nm-wavelength atomic transition. The latter is directed to either one of the two detection systems: the digital-inversion imaging and the optical-inversion detection, shown in Fig.~\ref{fig:setup}(b) and (c), respectively. 

For the initial characterization of the fiber, see Fig.~\ref{fig:setup}(a), we employ a digital micromirror device (DMD, Texas Instrument DLP4500NIR) to excite selected input modes of the fiber from its distal side, through the vacuum glass cell. The DMD is illuminated with a collimated Gaussian laser beam, the reflection of which is then expanded by a telescope (magnification $M=3$, not shown) and focused in the distal plane by a high-numerical-aperture aspheric condenser lens (focal length $f=26$ mm). Thus, contrary to usual DMD-systems relying on direct imaging with binary intensity modulation, we program binary holograms on the DMD in the Fourier plane to create arbitrary light patterns in the distal plane. The binary holograms are defined by switching on a pixel at position $\vec{r}$ on the DMD if $\cos \left(\phi\left(\vec{r}\right)\right) \geq 0$~\cite{Lee1979,Zupancic2016}, where $\phi\left(\vec{r}\right)$ is the desired phase profile after the reflection. We spatially control the first order generated by a grating-patterned hologram with a period $L_1 = 40$\!\unit{}{\micro\meter} ($\sim 8$ pixels) along the horizontal axis with unit vector $\vec{e}_{x}$. With the corresponding linear phase $\phi\left(\vec{r}\right) = \vec{r}.\vec{K}_1$ with $\vec{K}_1 = \vec{e}_{x} 2 \pi / L_1 $, the field in the distal plane is the series of focused beams of the different orders, generated by the grating pattern, with spacing $\lambda f / L_1 M \simeq 170$ \!\unit{}{\micro\meter}  for the atomic resonant wavelength $\lambda \simeq 780$ nm. This spacing being larger than the $100$-\!\unit{}{\micro\meter}  fiber radius, only the first order couples in the fiber at its center. From this point, we can displace this focused beam over a distance $\vec{q}$ in the distal plane by adding a linear phase $\vec{r}.\vec{q} \, 2 \pi M / \lambda f$.

The holographic configuration offers three main advantages over direct imaging. Firstly, the resolution of the position $\vec{q}$ of the beam is very precise, measured on the order of $10$ nm, because it results from the grating definition over the full DMD width. Secondly, the optical aberrations of the system can be measured and corrected efficiently~\cite{Zupancic2016}. To do so, we split the DMD in square zones of 19-pixel side and we apply the base linear-grating pattern only in a reference zone (center) and a measured zone. In the distal plane, we observe the resulting interference fringes whose phase at origin ($\vec{q} = 0$) corresponds to the wavefront distortion of the measured zone. As the aberrations vary with the beam's optical path, we perform this characterization for $9$ positions $\vec{q}_i$ roughly spanning the \unit{100 \times 100}{\micro\meter\squared} field of view, then interpolate the wavefront corrections for arbitrary positions $\vec{q}$. Despite the use of common lenses with aberrations and defects in DMD flatness, we thus achieve a Gaussian beam waist below \unit{1.2}{\micro\meter}  over the imaging region. Thirdly, arbitrary arrays of multiple focused spots at different positions are easily created by adding their respective holograms on the DMD, without observable interferences for spot-center distances larger than \unit{1.5}{\micro\meter} . We can also increase the diffraction-limited spot size by reducing the active area of the DMD, and displace the focal plane along the optical axis by adding a parabolic phase term ($\propto ||\vec{r}||^2$). We use these features of the DMD to generate micro-traps for atoms by DOPC~\cite{Vitrant2020} and to characterize the fiber transmission for imaging.  

As a first imaging method, we use a measurement of the optical field in the proximal space to reconstruct the field in the distal space, via digital inversion of the fiber's transmission matrix~\cite{Popoff2010}. We define and measure the transmission matrix $T$ on a basis of $10^4$ input modes $\{\vec{d}_i\}$ of \unit{1.2}{\micro\meter}  waist, created by the DMD setup. These modes are arranged in a planar square lattice of $100 \times 100$ spots with \unit{1}{\micro\meter} spacing at \unit{200}{\micro\meter} from the fiber's end face. The corresponding output modes $T \vec{d}_i$ are measured by off-axis holography~\cite{Cuche1999} where we combine on a beam splitter the output field and a coherent reference beam, with a small angle between their propagation axes. Thus, the measured light intensity on the camera presents angle-dependent interferences fringes. Taking digitally the Fourier transform of this image, we observe separated orders corresponding to multiples of the wavevector associated to the fringes. We isolate the first order and perform an inverse Fourier transform to recover the complex output field. 
For diffraction orders to be separated in Fourier space, the minimum speckle grain size $l_{\mathrm{spc}}$ must satisfy $l_{\mathrm{spc}} \geq 8 l_{\mathrm{pxl}}$ with $l_{\mathrm{pxl}}$ the pixel size of the camera. The magnification of the output field and the angle of the reference beam were optimized to reach this limit, which forces to spread the signal over many pixels. Nevertheless, off-axis holography allows us to measure a complex field with a single image. This feature has a particular importance in cold atoms setups where imaging is almost always a destructive operation and where atomic cloud preparation usually requires several seconds. 
For the output basis $\{\vec{p}_i\}$ of the proximal space that completes the definition of the fiber's transmission matrix, we select the $10^4$ pixels of the measured field that have the highest amplitude average over the input modes, offering the strongest signal. This square $10^4 \times 10^4$ transmission matrix $T$ is an empirical compromise between imaging quality and computer resources needed to store and invert images on the fly. We measure $T$ every week, a time span over which we observe only marginal degradation of the imaging quality. As experimental errors on small eigenvalues of $T$ have a large effect on $T^{-1}$, we use the mean square optimized inverting operator $W = (T^{\dagger}.T + \sigma I)^{-1}. T^{\dagger}$ with $\sigma$ the standard deviation of the experimental noise and $I$ the identity matrix~\cite{Popoff2010b}. Thus, from a measured field $\vec{p}$ in the proximal space, we get an image $W \vec{p}$ of the field in the distal space.  

\begin{figure}[t]
\centering
\includegraphics[width=85mm]{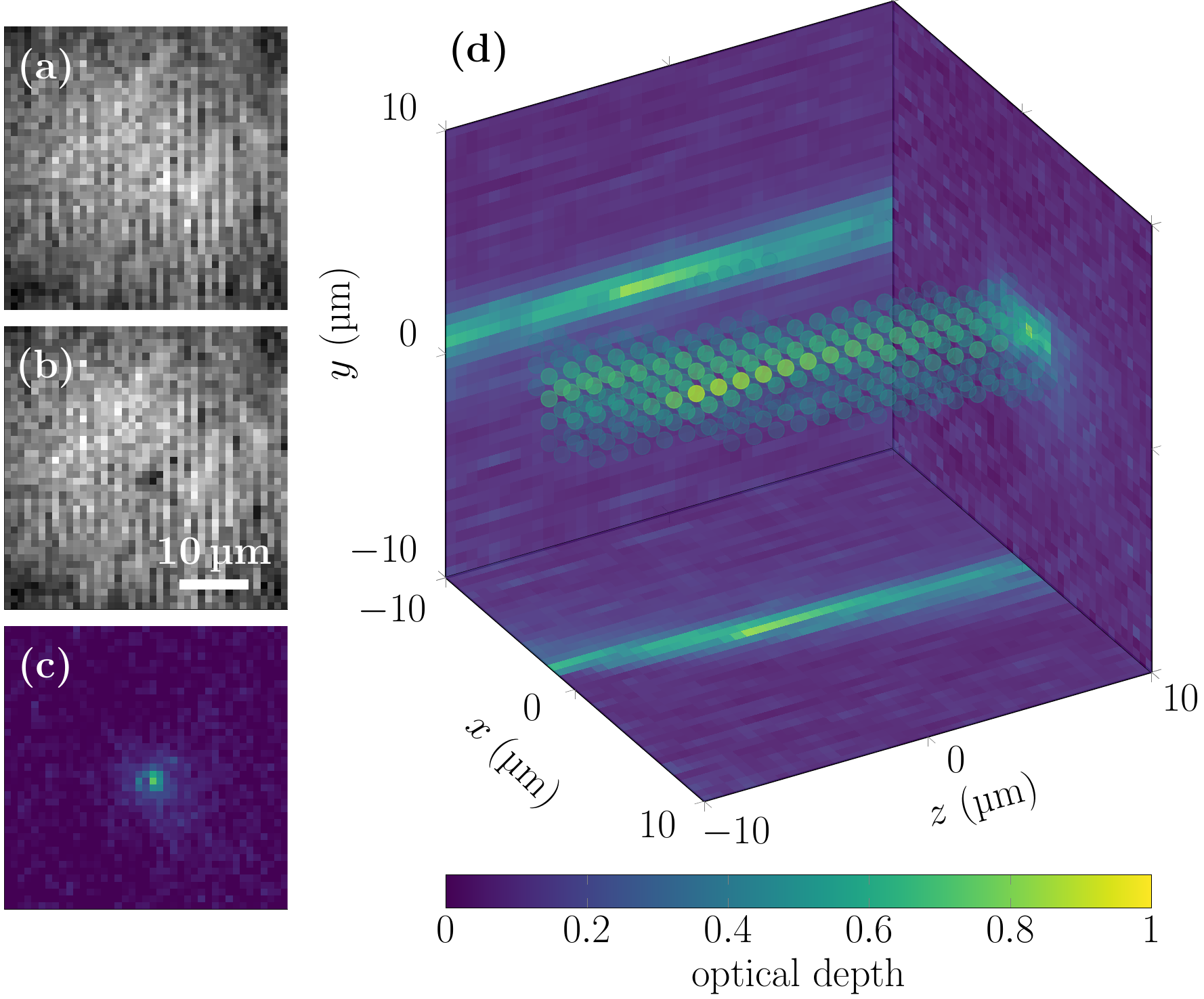}
\caption{ Absorption imaging through the multimode fiber by digital inversion. (a),(b) Light intensity in the distal plane, with (a) and without (b) atoms ($100$ averages). (c) Optical depth of the atomic cloud extracted from (a) and (b). (d) Optical depth in three dimensions reconstructed by digital inversion towards multiple distal planes separated by \unit{4}{\micro\meter}. Sides of the 3D-box show cuts through the point of maximum $OD$.}
\label{fig:DigInvIma}
\end{figure}  

We apply this digital-inversion technique to image cold atoms by absorption. While the reconstructed field intensity $I_{(a)}$ of a laser beam sent in the fiber shows a Gaussian profile (Fig.~\ref{fig:DigInvIma}(a)), a cold atomic cloud in a \unit{1.2}{\micro\meter}-waist microtrap casts a shadow at its center (Fig.~\ref{fig:DigInvIma}(b)). The corresponding optical depth ($OD = \log \left(I_{(a)}/I_{(b)}\right)$), shown on Fig.~\ref{fig:DigInvIma}(c), is proportional to the atomic density of the cloud. The observed sharp peak, with a Gaussian rms radius of \unit{1.7}{\micro\meter}, demonstrates the high resolution of digital-inversion imaging of cold atoms through the multimode fiber. We used this imaging to make temperature measurements with time of flight method or to determine the atomic distribution in multiple trap geometries~\cite{Vitrant2020}.

This holographic measurement of the field provides two additional advantages. Firstly, one could use it for phase contrast imaging, a powerful method to characterize very dense atomic clouds like Bose-Einstein condensates~\cite{Bradley1997}. Secondly, the field can be reconstructed for several distal planes from a single measurement by using the corresponding inverting matrices. We applied this three-dimensional imaging, shown on Fig.~\ref{fig:DigInvIma}(c), to our atomic cloud, where we observe the axial trapping length induced by the Rayleigh length of the trapping beam.  

The imaging by digital inversion is a powerful method for absorption imaging of dense atomic clouds, however it has two main limitations. Firstly, the interferometric imaging of the field required spreading the signal over the full camera sensor, and thus it has a much higher noise level than a single pixel detector. Secondly, the measurement relies on a coherent probe field and thus it cannot detect the fluorescence photons emitted by atoms with random phases. 

In order to overcome these limitations, we introduce another detection method: by optical inversion on a SLM, we send light coming from a chosen point in the object plane to a photon counter via a singlemode fiber (SMF), see Fig.~\ref{fig:setup}(c). We use DOPC with a setup similar to the one used to transport and trap the atoms~\cite{Vitrant2020}. During the characterization, for each pixel mode generated by the DMD in the distal plane, we measure the interference pattern between the speckle emerging at the proximal side and a reference beam from the SMF.  By using $3$-image phase-shifting interferometry, we determine the spatial phase pattern of the speckle relative to the SMF beam. To detect efficiently the photons emitted in the corresponding input mode, we apply the conjugate phase pattern on the SLM, which then couples the MMF output mode into the SMF input. We then connect the SMF ouput to an avalanche photodiode in photon counting mode.

\begin{figure}[t]
\centering
\includegraphics[width=85mm]{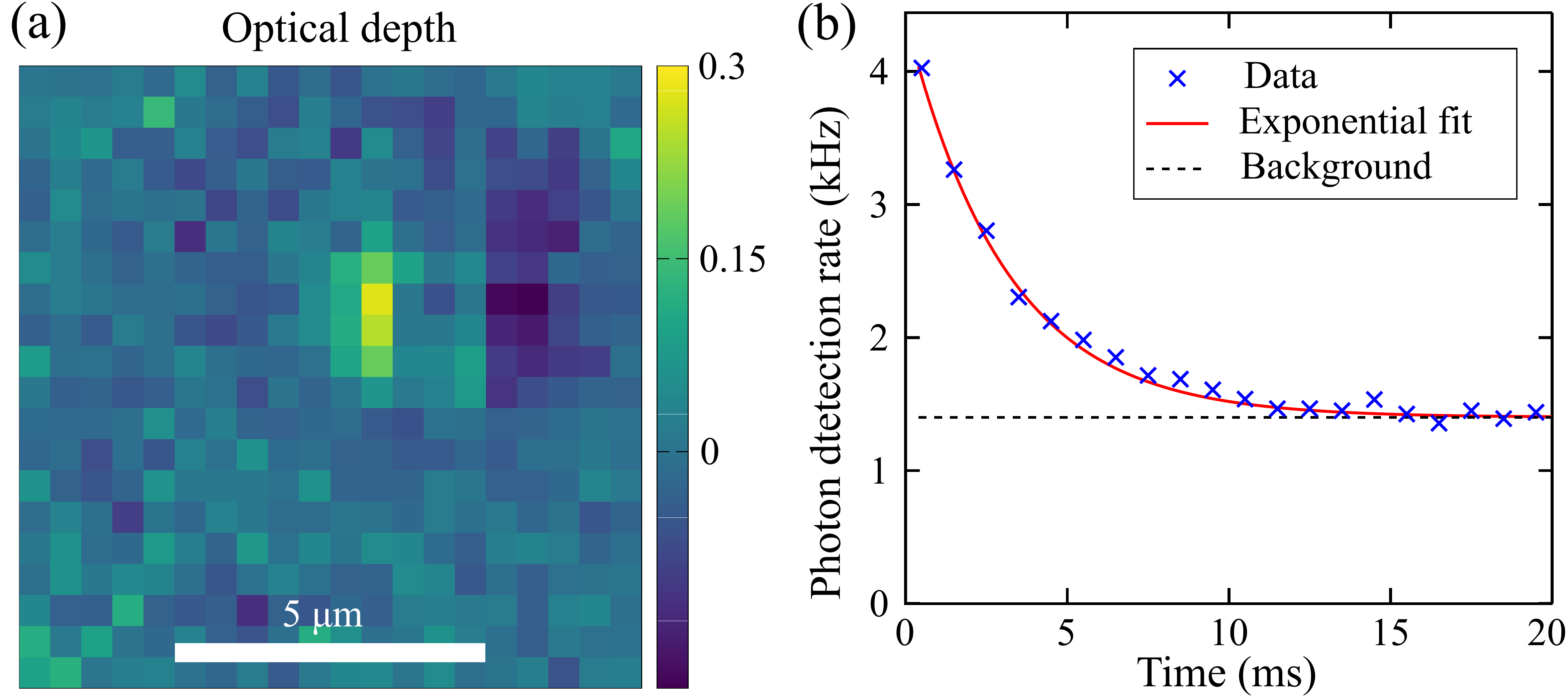}
\caption{ Detection through the multimode fiber by optical inversion. (a) Absorption observed pixel-by-pixel on the photon counter, while displacing the detected mode of the input plane with the SLM (averaged over $5$ repetitions per pixel). (b) Detected fluorescence on the photon counter as function of time after atom illumination starts (averaged over $10^4$ repetitions).  }
\label{fig:OptInvDet}
\end{figure}

As a first application, we use this optical-inversion detection to measure the absorption of atoms in a micro-trap. We change sequentially the SLM phase pattern to displace the detected mode in the distal plane with \unit{0.5}{\micro\meter} spacing. The resulting image of optical density, presented in Fig.~\ref{fig:OptInvDet}(a), shows a sharp peak corresponding to the center of the dipole trap, thus demonstrating a high resolution of this scanning-point imaging method. As expected, the noise is here significantly lower, with a standard deviation of $0.049(3)$ for 5 averages, than in the digital-inversion results, which requires $100$ averages to reach a similar standard deviation of $0.032(3)$. 

As a second application, we observe the incoherent fluorescence of atoms. We select the detection mode corresponding to the trap's center, where we observe the highest absorption. We illuminate the atoms with a resonant laser beam orthogonal to the fiber's axis (cf. Fig.~\ref{fig:setup}(c)) with an intensity nearly saturating the atomic transition. 
In Fig.~\ref{fig:OptInvDet}(b), the detected photon rate decays exponentially with a $3.1$-ms time constant. This represents the atom lifetime in the trap, dominated by heating from repeated absorption and emission of photons. 
The $1.4$-kHz background level stems from the spurious coupling to fiber modes of the illuminating beam ($1.1\,$kHz), the anti-Stokes Raman back-scattering~\cite{Farahani1999} in the MMF from the trapping light ($0.2\,$kHz) and the dark counts of the detector ($0.1\,$kHz). 
The level of Raman back-scattering is considerably reduced in comparison to atom micro-traps based on a singlemode fiber~\cite{Garcia2013}, thanks to a marginal overlap of trapping and detection lights in the fiber. 
The initial $2.6$-kHz detected photon rate above the background indicates the presence of approximately $3$ atoms on average in the micro-trap before the illumination starts. 

We deduce this atom number from the estimated total fluorescence detection efficiency $\eta_{\mathrm{tot}} \simeq 8 \cdot 10^{-5}$, which results from different contributions as $\eta_{\mathrm{tot}} = \eta_{\mathrm{col}} \eta_{\mathrm{SLM}} \eta_{\mathrm{opt}} \eta_{\mathrm{PC}}$. While the losses on standard optical elements ($\eta_{\mathrm{opt}} \simeq 0.9$, in particular on optical filters) and the photon counter efficiency ($\eta_{\mathrm{PC}} \simeq 0.7$) are not negligible, the dominant contributions are the efficiencies of the collection ($\eta_{\mathrm{col}} \simeq 1.3\%$) and of the SLM ($\eta_{\mathrm{SLM}} \simeq 1\%$). The collection efficiency is the overlap between the spontaneous emission mode of the atom and the $1.2$-\!\unit{}{\micro\meter}-waist detected mode. It has the same order of magnitude than those obtained for atomic micro-traps with aspheric lenses~\cite{Sortais2007, Garcia2013}. The SLM efficiency is an important limiting factor in our detection, but it also has a huge potential for improvements. First, the employed SLM (Hamamastu X10468-03) has $50\%$ absorption losses at $780\,$nm, while these losses can be reduced down to a few percent with other commercial models. Second, our system corrects only for a single polarization component of the speckle ($\sim 40\%$ losses) and full polarization control could be achieved by using two SLMs or two SLM sides~\cite{Moreno2012}. The third and most important contribution is the $\sim 4\%$ diffraction efficiency of the wavefront correction. Limited by our SLM resolution ($792\times 600$), this diffraction efficiency could be increased significantly towards the $\pi/4\simeq 79\%$ limit of phase modulation~\cite{Cizmar2011}, with a higher-resolution SLM ($4160 \times 2464$ available).  Such improvements would bring the detection method to the regime of single-atom single-shot detection. We are currently at the edge of this regime with a signal of approximately $2$ detected photons per emitting atom for $4$ photons of noise during the $3.1$-ms lifetime. 

The results presented here are a first attempt of using a multimode fiber to image cold atoms. Compared to traditional optical systems based on bulk aspheric lenses, it offers additional features such as the absence of a fixed working distance and a compactness that could be a game-changer for many experiments. Many of its drawbacks could be easily dealt with in future experiments, using modern hardware and eliminating initial mistakes. In particular, the fiber's transmission matrix can be easily made much more stable by eliminating or relocating the atomic dispensers, currently placed only a few mm away from the fiber and heated above \unit{500}{\degree}. Depending on the experimental requirements, a gradient-index fiber with a slightly lower numerical aperture could also offer a greater stability \cite{Caravaca-Aguirre2017}. Alternatively, one could use a shorter fiber allowing for a-priori modeling \cite{Ploeschner2015}, or imprint a partially-reflective pattern on its distal end to act as a permanently fixed calibration reference \cite{Gu2015}. Thus, we realistically expect that the need to optically access the distal end of the fiber for periodic recalibrations can be eliminated, making this optical system a viable and advantageous alternative to traditional ones, by allowing compact 3D imaging of cold atom clouds or single-shot detection of single atoms.

\section*{Acknowledgments}
 This work was funded by the DIM SIRTEQ project LECTRA, the IDEX grant ANR-10-IDEX-001-02-PSL PISE, and the ERC Starting Grant SEAQUEL. The authors thank P. Travers and  F. Moron for technical support, Q. Lavigne, T. Kouadou and J. Vaneecloo for their assistance at the early stage of the project, and S. Gigan for fruitful discussions.

\end{document}